\documentclass[conference]{./IEEEtran}
\IEEEoverridecommandlockouts
\usepackage{cite}
\usepackage{amsmath,amssymb,amsfonts, bm}
\usepackage{algorithmic}
\usepackage{graphicx}
\usepackage{epstopdf}
\graphicspath{{figures/}}
\usepackage{textcomp}
\usepackage{xcolor}
\usepackage{booktabs}
\usepackage{multirow}
\usepackage{subfigure}
\usepackage{stfloats}
\begin{document}

\title{Deep Modulation Recognition with Multiple Receive Antennas: An End-to-end Feature Learning Approach}

\author{\IEEEauthorblockN{ Lei Li, Qihang Peng, and Jun Wang}
\IEEEauthorblockA{\textit{University of Electronic Science and Technology of China, Chengdu, China}\\
\textit{Emails: anniepqh@uestc.edu.cn, junwang@uestc.edu.cn}}
}
\maketitle

\begin{abstract}
Modulation recognition using deep neural networks has shown promising advantages over conventional algorithms. However, most existing research focuses on single receive antenna. In this paper, two end-to-end feature learning deep architectures are introduced for modulation recognition with multiple receive antennas. The first is based on multi-view convolutional neural network by treating signals from different receive antennas as different views of a 3D object and designing the location and operation of view-pooling layer that are suitable for feature fusion of multi-antenna signals. Considering that the instantaneous SNRs could be different among receive antennas in wireless communications, we further propose weight-learning convolutional neural network which uses a weight-learning module to automatically learn the weights for feature combing of different receive antennas to perform end-to-end feature learning of multi-antenna signals. Results show that both end-to-end feature learning deep architectures outperform the existing algorithm, and the proposed weight-learning convolutional neural network achieves the best performance.
\end{abstract}


\section{Introduction}
Modulation recognition has been widely investigated over the past decades using likelihood-based (LB) or feature-based (FB) approaches\cite{nandi1998algorithms, wei2000maximum, yang1997suboptimal, swami2000hierarchical}. Inspired by the remarkable success of deep neural networks in computer vision and natural language processing\cite{krizhevsky2012imagenet}\cite{socher2014recursive}, modulation recognition using deep neural networks, also called deep modulation recognition, has shown promising performance improvements over conventional methods and attracted increasing research interests.

O'Shea first introduced convolutional neural networks (CNNs) for modulation recognition and showed performance advantages by learning features with deep neural networks over conventional expert features \cite{o2016convolutional}. In \cite{hong2017automatic}, recurrent neural network (RNN) was used to exploit temporal sequence characteristic of modulated signals.
Several deep architectures based on both CNN and RNN were examined and analyzed in \cite{west2017deep}. A CNN based unit classifier was proposed in \cite{meng2018automatic} to accommodate varying input lengths. In \cite{o2018over}, over-the-air radio machine learning dataset was generated using GNU Radio and USRP, showing the performance of deep modulation recognition under different channel conditions.
 In \cite{peng2018modulation}, the author provided a new way for CNN-based modulation recognition by representing received signal samples as images.
A two-stage hybrid method based on short-time Fourier transform (STFT) and CNN was proposed in \cite{daldal2020Automatic} which demonstrates the effectiveness of time-frequency features.
In \cite{shisheng2020deep}, the author introduced a deep neural network (DNN)-based modulation classifiers and shown its robustness to uncertain noise conditions.

However, most existing research focuses on single receive antenna. In radio propagation environments, modulation recognition signals may suffer from multipath fading,
which would cause performance degradations, especially when the signal experiences deep fade at the receiver\cite{o2018over}. One of the most powerful techniques to mitigate effects of fading is to use diversity-combining. Therefore, it is important to investigate deep neural networks for modulation recognition with multiple receive antennas.
In \cite{wang2020deep}, cooperative automatic modulation classification (Co-AMC) by multiple receive antennas is proposed for modulation recognition in MIMO systems, where the CNN is trained with each receive antenna independently, and the final decision on modulation type is reached by cooperative decision. Results show that Co-AMC by multiple receive antennas performs better than single receive antenna. However, the performance of this method is limited by feature learning with single-antenna signals.

In this paper, two end-to-end feature learning deep architectures, which can be trained directly using multi-antenna signals, are introduced for modulation recognition with multiple receive antennas, including:

\begin{itemize}
\item Multi-View CNN (MVCNN). This was initially proposed for 3D shape recognition using view-based descriptors of 2D images, which combines information from multiple views of a 3D shape into a single and compact descriptor for end-to-end feature learning\cite{su2015multi}. By treating signals from different receive antennas as different views of a 3D object and designing a suitable view-pooling location
and operation for feature fusion of multi-antenna signals, MVCNN can then be applied for modulation recognition with multiple receive antennas.

\item Weight-Learning CNN (WLCNN). Considering that in wireless communications, the instantaneous signal-to-noise ratio (SNR) of signals could be different among receive antennas, the feature combing weight of each antenna need to be adaptive according to input signal characteristics. We propose WLCNN based on a weight-learning module (WLM) which is designed to  automatically learn the weights for combining features of different receive antennas, and incorporated with CNNs to enable end-to-end feature learning for modulation recognition with multiple receive antennas.


\end{itemize}


Results show that comparing with existing Co-AMC method, both end-to-end feature learning deep architectures, MVCNN and WLCNN, gain better modulation classification accuracy. Further, the proposed WLCNN, which automatically learns the weights for feature combing, results in the best performance.

The rest of this paper is organized as follows. System model is given in Section \uppercase\expandafter{\romannumeral2}, and deep architectures for multi-antenna modulation recognition are described in Section \uppercase\expandafter{\romannumeral3}. In Section \uppercase\expandafter{\romannumeral4}, simulation results are given and analyzed. Finally, conclusions are discussed in Section V.

\section{System Model}
\subsection{Signal Model and Representations}

Suppose that signals are sent from a transmitter and experience flat-fading in radio channels, our goal is to determine the modulation type of the signals using a receiver with $N_r$ antennas. Given the transmitted signal $\mathbf{s}$, the equivalent baseband received signal $\mathbf{r}$ can be expressed as:

\begin{equation}
\mathbf{r}=\mathbf{H} \mathbf{s}+\mathbf{n}
\end{equation}

\noindent where $\mathbf{H}$ is a complex-valued $N_r \times 1$ vector where its $n_r$-th ($1\leq n_r\leq N_r$) element denotes the complex channel coefficient between the transmitter and the $n_r$-th receiving antenna. $\mathbf{n}$ denotes the additive Gaussian noise.

Within one observation interval, $N$ signal samples are collected from each antenna to form a $1 \times N$ complex-valued vector, which is further decomposed into a $2 \times N$ matrix, where the first and second row correspond to the in-phase and quadrature components, respectively. Signal samples from $N_r$ receive antennas are then collected to form a $N_r \times 2 \times N$ tensor. For more efficient training, we normalize the power of these vectors.

Let $\bold{x}^{(i)}$ denote the three-dimensional tensor collected in the $i$-th observation interval, and $\bold{y}^{(i)}$ be its corresponding label denoting the transmitted signal's modulation type. Then $\big(\bold{x}^{(i)}, \bold{y}^{(i)} \big)$ forms one training example, and training examples from randomly different time instants with identical observation durations are collected as datasets for deep modulation recognition.

\subsection{Deep Learning Based Modulation Recognition}
The optimization of supervised deep learning process can be formulated as finding the parameters $\boldsymbol{\theta}$ of a neural network that significantly reduce a cost function $J(\boldsymbol{\theta})$\cite{goodfellow2016deep}, which can be written as,
\begin{equation}
J(\boldsymbol{\theta})=\mathbb{E}_{(\bold{x}, \bold{y}) \sim \hat{p}_{\text {data }}} L(f(\bold{x} ; \boldsymbol{\theta}), \bold{y})
\end{equation}

\noindent where $L(\cdot)$ denotes the loss function, $f(\bold{x} ; \boldsymbol{\theta})$ corresponds to the neural network predicted output when the input signal is $\bold{x}$, and $y$ denotes the true modulation class label, $\hat{p}_{\text {data }}$ is the empirical distribution of modulated signals.

In our experiments, the cross-entropy is chosen as the loss function, given by
\begin{equation}
L(\hat{y}, y)= - \sum_{m=1}^{M}y_m\log(\hat{y}_m)
\end{equation}

\noindent where $M$ is the total of modulation types.

For the training of neural network parameters, Adam optimizer\cite{kingma2014adam} with initial learning rate 0.001 is used for mini-batch optimization.

\section{Deep Architectures for Multi-Antenna Modulation Recognition}
In this section, two end-to-end feature learning deep architectures including MVCNN and WLCNN, which can be trained directly using multiple antenna signals, are introduced for modulation recognition with multiple receive antennas, Co-AMC \cite{wang2020deep} is also introduced for comparison.

 \begin{figure}[!t]
\centering
\includegraphics[width=1.3in]{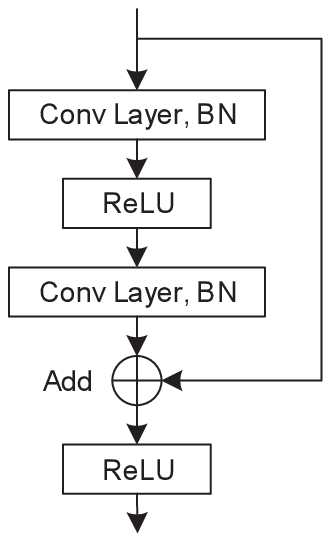}
\caption{The architecture of a residual block.}
 \label{fig:resBlock}
 \end{figure}

\begin{table}[!t]
\renewcommand{\arraystretch}{1.3}
\caption{Base CNN LAYOUT}
\label{tab:ResNet34}
\centering
\begin{tabular}{llccc}
\toprule
                                                    & Layer                                      &Kernel Size    &Stride    &Output Size   \\
 \midrule
                                                    & Input                                         &                   &              &$1 \times 2 \times 512$   \\
 \multirow{3}{*}{CNN$_1$} & Conv, BN, Tanh                       &$7$           &2             &$64 \times 1 \times 256$  \\
                                                & Max Pooling                           &$3$          &2              &$64 \times 1 \times 128$  \\
                                                & Residual Block * 3                 &$3$         &1               &$64 \times 1 \times 128$  \\
 \midrule
 \multirow{5}{*}{CNN$_2$}  & Residual Block * 4                 &$3$         &2               &$128 \times 1 \times 64$  \\
                                                 & Residual Block * 6                 &$3$         &2                &$256 \times 1 \times 32$  \\
                                                & Residual Block * 3                 &$3$         &2                &$512 \times 1 \times 16$  \\
                                                & Average Pooling                  &$16$        &1                &$512 \times 1 \times 1$  \\
                                                & FC, Softmax                           &                   &                &$ 20$             \\
\bottomrule
\end{tabular}
\end{table}

ResNet\cite{he2016deep} is chosen as a base network to construct the multi-antenna modulation recognition architectures. This is because: (1) ResNet can reduce the effect of degradation of deeper networks; (2) Existing research on modulation recognition with single receive antenna has shown that ResNet is a suitable candidate\cite{o2018over}. The ResNet is tuned such that when we further increase the network size or change network parameters, the modulation classification performance does not further improve, given the single-antenna dataset. In this way, a 34-layer ResNet is obtained and shown in Table \ref{tab:ResNet34}, which consists of 1 convolutional (Conv) layer, 16 residual blocks, and a fully-connected (FC) layer, the total parameters of the base CNN is about 7.23M. Each residual block consists of two convolutional layers and batch normalization (BN)\cite{ioffe2015batch} operations, as illustrated in Fig. \ref{fig:resBlock}. Considering that the modulated signals are composed of both positive and negative values, Tanh is used as the activation function on first convolutional layer and ReLU on the others, and Softmax function is used to normalize the output distributions.

\subsection{Multi-View Convolutional Neural Network}
MVCNN was initially proposed for 3D view-based shape recognition, which combines information from multiple views of a 3D shape into a single and compact shape descriptor for end-to-end feature learning and has shown to be quite effective in 3D shape recognition.

\begin{figure}[!t]
\centering
\includegraphics[height=1.3in]{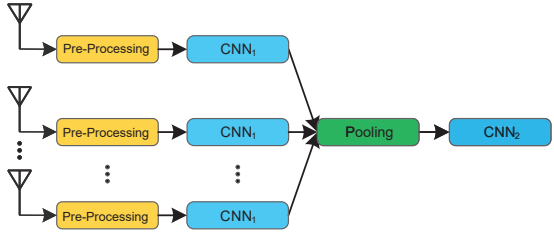}
\caption{The architecture of MVCNN for multiple-antenna modulation recognition.}
 \label{fig:mvcnn}
 \end{figure}

To realize the MVCNN in our considered scenario, the received signals from one antenna is taken as one view of 3D object, and the base network, i.e., the ResNet in Table \ref{tab:ResNet34}, is split by a view-pooling layer into two parts: CNN$_1$ and CNN$_2$. As illustrated in Fig. \ref{fig:mvcnn}, features from individual antennas are first extracted by CNN$_1$, and then fused by view-pooling layer. Note that all the $N_r$ branches of MVCNN, i.e., CNN$_1$, share the same parameters. Operations of the view-pooling layer are similar to conventional pooling layers in CNN, e.g., max pooling or average pooling, the difference lies in that the view-pooling operations are carried out across the dimension of receive antennas. Features fused by view-pooling layer are then passed through CNN$_2$, where the information obtained across multiple antennas is further processed and then the output of MVCNN is obtained, which corresponds to a vector consisting of the empirical conditional probabilities of different modulation types.

Different locations of the view-pooling layer determine different network architectures and could lead to different modulation recognition performances. So the location of view-pooling layer and pooling operations need to be carefully designed. In this work, we test the performance of MVCNN with different locations and operations, including maximum and mean operations of the view-pooling layer. Results show that the best performance is obtained by locating the view-pooling layer after the first 3 residual blocks, and by using max-view-pooling operation for feature fusion across the dimension of receive antennas, where the max-view-pooling is an element-wise maximum operation across different receive antennas as:
\begin{equation}
F_{o}=\max\{F_1, F_2, ..., F_{N_r}\}
\label{eq:wb}
\end{equation}
\noindent where $F_{nr} (nr=1, 2, ..., N_r)$ denotes signal features from the $nr$-th antenna, and $F_o$  is the output feature of max-view-pooling.

\subsection{Weight-Learning Convolutional Neural Network}
In wireless communications, radio signals received from different receive antennas could experience different fading, and the weight of feature combing for each antenna need be adaptive to the input signal characteristics. We propose WLCNN for automatically learning the weights for feature combing from different receive antennas.

\begin{figure}[!t]
\centering
\includegraphics[height=1.3in]{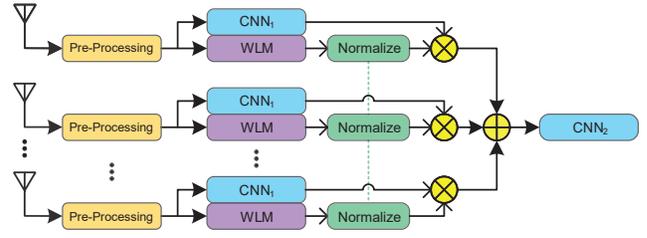}
\caption{The architecture of WLCNN for multiple-antenna modulation recognition.}
 \label{fig:wlcnn}
 \end{figure}

The architecture of WLCNN is shown in Fig. \ref{fig:wlcnn}, the base CNN, as Table \ref{tab:ResNet34}, is split into two parts: CNN$_1$ and CNN$_2$. Features from individual antennas are first extracted by CNN$_1$, then the signal features from  different receive antennas are combined with learned weights from the WLM. Combined features are then passed through CNN$_2$ to reach the output of WLCNN. Similar as that for  MVCNN, the location of feature combing is located after the first 3 residual blocks, and CNN$_1$ share the same parameters.

The process of learning the weights of signal feature combing from different receive antennas is described as follows.

Given the signal feature from the $nr$-th receive antenna $F_{nr}$, the combined feature $F_o$ can be written as an element-wise weighted sum of each receive antenna, given by
\begin{equation}
F_{o}=\sum_{nr=1}^{N_r}w_{nr}*F_{nr}
\label{eq:wb}
\end{equation}
\noindent where ${w}_{nr}$ is the normalized learned weight of the $nr$-th receive antenna.

Due to the unknown characteristics of received signals, a neural network based WLM is adopted to learn the weight for each receive antenna.
Specifically, for the input signal from the $nr$-th receive antenna $x_{nr}$, the WLM with parameters $\boldsymbol{\theta_{w}}$ outputs a single value $\hat{w}_{nr}$ representing the estimated weight of the $nr$-th receive antenna, which can be expressed  as,
\begin{equation}
\hat{w}_{nr}=g(x_{nr}; \boldsymbol{\theta_{w}})
\label{eq:wl}
\end{equation}

The estimated weights of different receive antennas are normalized by softmax function as,
\begin{equation}
w_{nr}=\frac{e^{\hat{w}_{nr}}}{\sum_{nr=1}^{N_r}e^{\hat{w}_{nr}}}
\label{eq:softmaxnorm}
\end{equation}

The WLM is designed as a CNN based neural network with only one neuron in the last layer representing the weight of feature combing, and this weight automatically varies with different input signals. The WLM for each receive antenna shares the same structure and is composed of 1 convolutional layer and 3 residual blocks with 16 convolutional filters. Similar to  MVCNN, the parameters of WLCNN are trained in an end-to-end manner with multi-antenna signals. The total parameters of the WLM is around 6.34k and is roughly 0.1\% parameters of base CNN. The specific architecture of the WLM is as Table \ref{tab:wlm}.
\begin{table}[!t]
\renewcommand{\arraystretch}{1.3}
\caption{Architecture of WLM}
\label{tab:wlm}
\centering
\begin{tabular}{@{}llcccl@{}}
\toprule
 & Layer                                      &Kernel Size    &Stride    &Output Size   \\
 \midrule
 & Input                                         &                   &              &$1 \times 2 \times 512$   \\
 & Conv, BN, Tanh                     &$7$           &2             &$16\times 1 \times 256$  \\
 & Residual Block                      &$3$           &2             &$16 \times 1 \times 128$  \\
 & Residual Block                      &$3$           &2             &$16 \times 1 \times 64$  \\
 & Residual Block                      &$3$           &2             &$16 \times 1 \times 32$  \\
 & FC                                             &                &               &$1$             \\
\bottomrule
\end{tabular}
\end{table}

\subsection{Cooperative automatic modulation classification}
\begin{figure}[!t]
\centering
\includegraphics[width=2.8in]{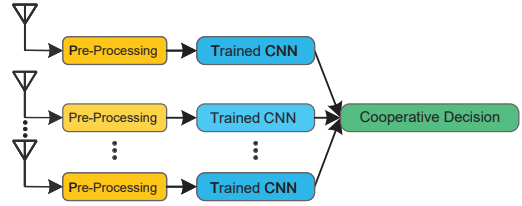}
\caption{The decision process of cooperative automatic modulation classification (Co-AMC) for multiple-antenna modulation recognition.}
 \label{fig:egcnn}
 \end{figure}

In Co-AMC, a base CNN is first trained by signals from each receive antenna, and then the decision on the modulation type is cooperatively made based on outputs of trained CNNs from all branches. Direct averaging (DA) method is used due to the identical signal distribution of each receive antenna in our experiment, where each branch of predicted distribution is averaged to make a cooperative prediction as Fig. \ref{fig:egcnn}. Also, these CNNs share the same parameters and have the same structure.

The output of the $k$-th CNN is a $M \times 1$ vector denoted as $\bold{\hat{p}_k}$, where $M$ corresponds to the total number of modulation types, and its $m$-th element, $\hat{p}_{km}$, can be seen as the predicted possibility of the $m$-th modulation type from the $k$-th receive antenna.

The predicted distributions from different antennas are averaged to obtain the global estimate of modulation type. Let $\hat{p}^{(m)}$ denote the estimated probability of the $m$-th modulation format, given by
\begin{equation}
\hat{p}^{(m)}=\sum_{k=1}^{N_{r}} \hat{p}_{k m} / N_{r}
\label{eq:egc}
\end{equation}

The decision on the modulation format can be reached by choosing the index $m^*$ that maximizes $\hat{p}^{(m)}$, given by
\begin{equation}
 m^* = \mathop{\arg\max }_{m}\hat{p}^{(m)}
 \label{eq:egcdecision}
\end{equation}

\section{Results and Analysis}
In this section, we present the performances of end-to-end feature learning deep architectures and compare with the Co-AMC for modulation recognition with multiple receive antennas.

\begin{figure}[!t]
\centering
\subfigure[MVCNN]{\includegraphics[width=2.95in]{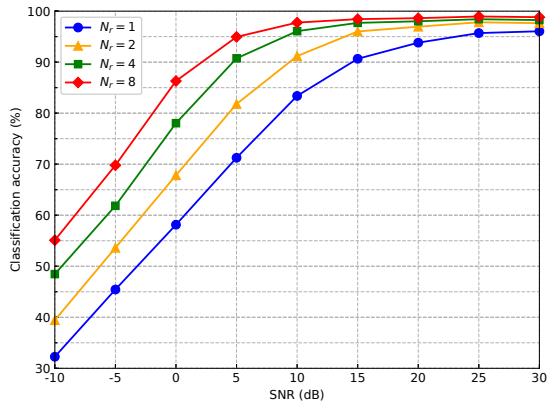}}
\vfil
\subfigure[WLCNN]{\includegraphics[width=2.95in]{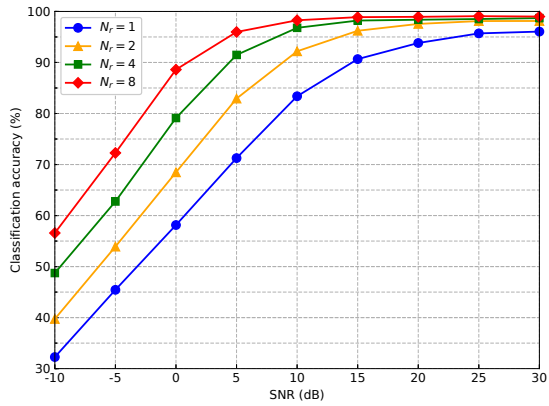}}
\vfil
\subfigure[Co-AMC]{\includegraphics[width=2.95in]{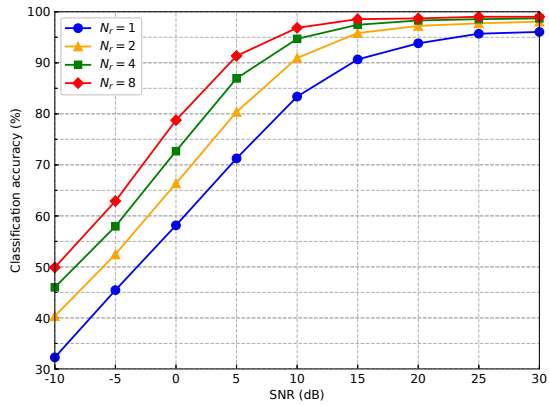}}

\caption{Modulation recognition performances versus SNR with different numbers of receive antennas with different deep architectures.}
\label{fig:acc_snr_model}
 \end{figure}

Datasets in our experiments are generated using GNU Radio\cite{o2016radio}. A square root raised cosine filter with a roll-off factor of 0.35 is used for pulse shaping. Transmitted signals experience independent and identically distributed Rayleigh fading, where the second moment of the Rayleigh fading coefficient is normalized to unit. Received signals are filtered and down converted to baseband, and are up sampled by a factor of 8. $N_r \times 2 \times N$ real samples are collected from $N_r $ receive antennas to form one example, where $N_r$ ranges from 1 to 8 and $N$ is set to 512 in our experiments. 2000 examples are generated for each SNR and each modulation type, and the size of training set and test set is 1 : 1. There are 20 different modulation formats including both analog and digital modulation types, including: BPSK, QPSK, 8PSK, 16PSK, 16QAM, 32QAM, 64QAM, 128QAM, 256QAM, 16APSK, 32APSK, 64APSK, 128APSK, OOK, 4ASK, GMSK, FM, AM, DSB, SSB.

  \begin{figure}[!t]
\centering
\includegraphics[width=3in]{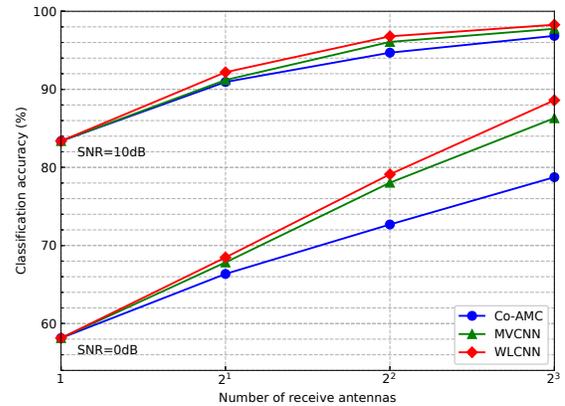}
\caption{The classification accuracy comparison of different multi-antenna modulation recognition methods versus number of receive antennas with SNR = 0dB and 10dB.}
 \label{fig:acc_nr_model}
 \end{figure}

The classification accuracy versus SNR for MVCNN, WLCNN and Co-AMC with different numbers of receive antennas are presented in Fig. \ref{fig:acc_snr_model}.
It is shown that, when the number of receive antennas increases, the classification accuracies increase accordingly.
When $Nr$ increases from 1 to 8, the noise tolerance of the MVCNN and WLCNN is improved by around 10 dB, and the classification accuracy is improved by up to 30\% when the SNR is about 0dB.
 This coincides with the analysis in conventional modulation recognition for multiple receive antennas that recognition performance can be improved by utilizing spatial diversity. Note that performances of the multi-antenna modulation recognition methods are identical for the case $N_r=1$. This is because when the number of receive antenna is equal to 1, these networks reduce to the same computational architecture.

The classification accuracy versus $N_r$ for different architectures with given SNRs are plotted in Fig. \ref{fig:acc_nr_model}. It is shown that comparing with Co-AMC,  the two end-to-end feature learning deep architectures have better performance. When the $Nr$ increases to 8, MVCNN improves the modulation classification accuracy by about 7.4\% while WLCNN improves by about 9.8\% in 0dB, this is due to the advantages of end-to-end feature learning with radio signals from multiple antennas simultaneously. The proposed WLCNN results in the best performance among them, this is because, although the channel coefficients are independent and identically distributed, at any particular instant of time, the fading coefficient for one channel is different from another. That is, the instantaneous SNR from different receive antennas are different. The proposed WLCNN automatically learns the weights for feature combining of different antennas with a WLM. In this way, the weights for multiple receive antennas are adaptively varied according to different input signals. In other words, the proposed WLCNN, which learns the weights for feature combing through end-to-end training, can better fuse features from multiple receive antennas than MVCNN which uses a pooling operation.

\section{Conclusion and Discussion}
In this paper, two end-to-end learning deep architectures are introduced for modulation recognition with multiple receive antennas: MVCNN and WLCNN. Compared with existing Co-AMC algorithm, the two end-to-end deep architectures for modulation recognition with multiple receive antennas achieve better performance. Further, the proposed WLCNN obtains the best performance among them by automatically learning the feature combining weights of different antennas with adapting to different input signals.

\section{Acknowledgment}
We would like to thank Professor Nuno Vasconcelos for his valuable comments and discussions on deep architectures for modulation recognition.

\bibliography{reference}

\end{document}